\documentclass[runningheads]{svmult}
\usepackage{graphicx}
\usepackage{latexsym}
\usepackage{subeqnar}
\begin{document}
\title*{The Inelastic Maxwell Model}
\toctitle{The Inelastic Maxwell Model}
\titlerunning{The Inelastic Maxwell Model}
\author{E.~Ben-Naim\inst{1}\and P.~L.~Krapivsky\inst{2}}
\authorrunning{Ben-Naim and Krapivsky}
\institute{Theoretical Division and Center for Nonlinear Studies,
Los Alamos National Laboratory, Los Alamos, NM 87545 \and Center
for Polymer Studies and Department of Physics, Boston University,
Boston, MA 02215} \maketitle
\begin{abstract}

Dynamics of inelastic gases are studied within the framework of
random collision processes.  The corresponding Boltzmann equation
with uniform collision rates is solved analytically for gases,
impurities, and mixtures. Generally, the energy dissipation leads
to a significant departure from the elastic case. Specifically,
the velocity distributions have overpopulated high energy tails
and different velocity components are correlated. In the freely
cooling case, the velocity distribution develops an algebraic
high-energy tail, with an exponent that depends sensitively on
the dimension and the degree of dissipation. Moments of the
velocity distribution exhibit multiscaling asymptotic behavior,
and the autocorrelation function decays algebraically with time.
In the forced case, the steady state velocity distribution decays
exponentially at large velocities. An impurity immersed in a
uniform inelastic gas may or may not mimic the behavior of the
background, and the departure from the background behavior is
characterized by a series of phase transitions.

\end{abstract}

\section{Introduction}

Granular gases consist of weakly deformable macroscopic particles
that interact via contact interactions and dissipate energy during
collisions \cite{pl}. In typical experimental situations the
system is constantly supplied with energy to counter the loss
occurring during inelastic collisions \cite{jnb}.  The most
ubiquitous features of granular gases are their tendency to form
clusters and their anomalous velocity statistics
\cite{lcdkg,ou1,rm,wp}. In turn, these characteristics lead to
many interesting collective phenomena including density
inhomogeneities \cite {kwg}, shocks \cite{slk,rbss}, size
segregation \cite{hk}, pattern formation \cite{mus}, and phase
transitions \cite{ou}, to name a few.

Inelastic hard spheres provide a useful theoretical,
computational, and experimental framework for studying granular
gases \cite{h,mac,gzb,bssms}. Inelastic gases pose new theoretical
challenges as their dynamics are governed by dissipative
collisions and strong velocity correlations. Fundamental
equilibrium characteristics such as ergodicity and equipartition
of energy are typically violated by this nonequilibrium gas
system. As a result, inelastic gases exhibit finite time
singularities \cite{bm,my,zk}, chaotic behavior \cite{dlk},
breakdown of molecular chaos \cite{gz,be,cornell,bcdr,nbc1}, and
anomalous velocity statistics \cite{ep,db,van,plmv,bbrtv}.

The significance of the energy dissipation is nicely demonstrated
by considering a mean-field version of the hard sphere system,
namely a random collision process. In this process, randomly
chosen pairs of particles undergo inelastic collisions with a
random impact direction. This process, often called the Maxwell
model, is described by a Boltzmann equation with a velocity
independent collision rate. As this model is random in all
respects, it is analytically tractable. In fact, in classical
kinetic theory of gases, the Maxwell model precedes the Boltzmann
equation \cite{max}.  Historically, it played an important role in
the development of kinetic theory
\cite{classical,true,krupp,e,bob}, and it remains the
subject of current research \cite{bag,bobyl,hayak}.  Recently, it has
been realized that the Maxwell model is analytically tractable
even for inelastic collisions
\cite{bk,bcc,bmp,kb,bk1,bk2,eb1,eb2,erb,bc,mp,mp2,droz,santos,hh,vg}.
In this review, we detail dynamics of uniform inelastic gases,
isolated impurities in uniform gases, and binary mixtures.

First, we consider the one-dimensional gas where explicit
analytic solutions are possible (Sec.~2).  In the freely cooling
case, the velocity distribution approaches a universal
(dissipation independent) scaling form with an algebraic
high-energy tail. Furthermore, the moments of the velocity
distribution exhibit a multiscaling asymptotic behavior, and the
velocity autocorrelation function decays algebraically with a
non-universal exponent.  In the uniformly heated case, the system
approaches a steady state, and the velocity distribution has an
exponential high-energy tail.

Many of these features extend to higher dimensions (Sec.~3). In
the freely cooling case, however, the scaling function underlying
the velocity distribution is no longer universal --- its shape and
its extremal behavior depend on the spatial dimension and the
degree of dissipation.  We obtain explicit expressions for low
order moments and show the development of correlations between
different velocity components. Such correlations are more
pronounced in the freely cooling case, compared with the forced
case.

When an impurity is immersed in a homogeneous cooling fluid, various
scenarios are possible (Sec.~4). If the impurity mass is lighter than
a critical mass, its velocity statistics are completely governed by
the background. As the impurity mass increases, velocity moments of
higher order may diverge asymptotically compared with the fluid. A
series of critical masses govern these phase transitions.  Ultimately,
the temperature ratio may diverge, and the impurity becomes so
energetic that it effectively decouples from the fluid.  In this case,
its velocity distribution is distinct from the fluid.  Although
single-time velocity statistics of the impurity may be governed by the
fluid, two-time statistics such as the autocorrelation function are
always different.  

For mixtures (Sec.~5), all components have the same asymptotic
temperature decay, and a high-energy tail as in the uniform gas case.
This behavior is shown in detail for one-dimensional freely cooling
binary mixtures.

We finally describe a lattice generalization of the Maxwell model
where particles are placed on a lattice and only nearest neighbors
interact via inelastic collisions (Sec.~6). We show the
development of spatial velocity correlations with a diffusively
growing correlation length.  These correlations significantly
reduce the temperature cooling rate.

\section{Uniform Gases: One Dimension}

\subsection{The freely cooling case}

Consider an infinite system of identical particles that interact
via instantaneous binary collisions. When a pair of particles with
velocities $(u_1,u_2)$ collide, their post collision velocities
$(v_1,v_2)$ are given by the collision rule
\begin{equation}
\label{rule-1d} v_{1,2}=u_{1,2}\mp(1-p)g,
\end{equation}
with $g=u_1-u_2$ the relative velocity, and $p$ the collision
parameter.  The total momentum is always conserved,
$u_1+u_2=v_1+v_2$. Since the system is invariant under the
Galilean transformation $v\to v-v_0$, we set the average velocity
to zero, without loss of generality. In each collision, the
relative velocity between the colliding particles is reduced by
the restitution coefficient $r=1-2p$.  In an inelastic collision,
there is an energy loss, $\Delta E=-p(1-p)g^2$ (particle mass is
set to unity); energy loss is maximal for completely inelastic
collisions ($p=1/2$), and it vanishes for elastic collisions
($p=0$).

In a random collision process, the collision rate is independent
of the particles' velocities. Let $P(v,t)$ be the normalized
velocity distribution at time $t$. It evolves according to the
Boltzmann equation
\begin{eqnarray}
\label{be-1d} {\partial P(v,t)\over\partial t}=&K&
\int du_1\, P(u_1,t)\int du_2\,P(u_2,t)\\
&\times&\left\{\delta\left[v-u_1+(1-p)g\right]-\delta(v-u_2)\right\}.\nonumber
\end{eqnarray}
This Boltzmann equation, with a velocity independent collision
kernel, is termed the Maxwell model in kinetic theory \cite{e}.
The overall collision rate $K=\sqrt{T}$, with the granular
temperature $T=\int dv\, v^2 P(v,t)$, is chosen to represent the
typical particle velocity.  The quadratic integrand in
Eq.~(\ref{be-1d}) reflects the random and binary nature of the
collision process, while the gain and loss terms reflect the
collision rules (\ref{rule-1d}). One can verify that the total
number of particles, $\int dv P(v,t)$, and the total momentum,
$\int dv \, v P(v,t)$, are conserved; hereinafter we set $\int dv
P(v,t)=1$ and $\int dv \, v P(v,t)=0$.

For inelastic hard spheres, where the collision rate equals the
relative velocity, the rate equation governing the temperature is
part of an infinite hierarchy of equations. In contrast, in the
Maxwell model, the temperature obeys a closed rate equation
\begin{equation}
\label{temp-eq-1d} {d T\over dt}=-\lambda T^{3/2}\qquad {\rm
with}\qquad \lambda=2p(1-p).
\end{equation}
The temperature decays algebraically with time according to Haff's
law \cite{pkh}
\begin{equation}
\label{temp-1d} T(t)={T_0\over (1+t/t_0)^2},
\end{equation}
with the time scale $t_0^{-1}=p(1-p)\sqrt{T_0}$. Eventually, all
particles acquire the average velocity: $P(v,\infty)=\delta(v)$.
Our goal is to characterize how the velocity distribution
approaches this terminal state asymptotically. Since the overall
collision rate is uniform, it is useful to characterize time by
the collision counter $\tau=\int_0^t dt' K(t')$, equal to the
average number of collisions experienced by a particle,
\begin{equation}
\label{tau-1d} \tau=\frac{2}{\lambda}\,\ln (1+t/t_0).
\end{equation}
In terms of the number of collisions,
$T(\tau)=T_0\,e^{-\lambda\tau}$.

The convolution structure of the Boltzmann equation suggests to
apply the Fourier transform $F(k,\tau)=\int dv\, e^{ikv}
P(v,\tau)$. This quantity evolves according to \cite{bk}
\begin{equation}
\label{four-eq-1d} {\partial
\over\partial\tau}\,F(k,\tau)+F(k,\tau)=F(k-pk,\tau)\,F(pk,\tau) .
\end{equation}
This closed equation is both nonlinear and nonlocal, yet it is
analytically tractable.  Assuming that the velocity distribution
approaches its final state in a self-similar fashion, we seek a
scaling solution
\begin{equation}
\label{dis-scl-def-1d} P(v,t)=T^{-1/2}{\cal P}(w) \qquad {\rm
with}\qquad w=v\,T^{-1/2}.
\end{equation}
The scaling form corresponding to the Fourier transform is
$F(k,\tau)=f(z)$ with the variable $z=|k|\,T^{1/2}$. Substituting
the scaling ansatz into Eq.~(\ref{four-eq-1d}) and using
$dT/d\tau=-\lambda T$, the Fourier scaling function satisfies
\begin{equation}
\label{four-scl-eq-1d} -p(1-p)f'(z)+f(z)=f(z-pz)f(pz).
\end{equation}
This equation is supplemented by the small-$z$ behavior $f(z)\cong
1-{1\over 2}z^2$, which is dictated by the small wave number
behavior, $F(k)\cong 1-{1\over 2}k^2T$. Subject to these
conditions, the (unique) solution is \cite{bmp}
\begin{equation}
\label{four-scl-1d} f(z)=\left(1+z\right)e^{-z}.
\end{equation}
The scaled velocity distribution is obtained by performing an
inverse Fourier transform
\begin{equation}
\label{dis-scl-1d} {\cal P}(w)={2\over \pi}\,{1\over (1+w^2)^2}.
\end{equation}

Remarkably, the scaled velocity distribution is independent of the
collision parameter $p$. Another important feature is the
algebraic decay: ${\cal P}(w)\sim w^{-4}$ for $w\gg 1$.  This
behavior should be compared with the exponential high energy tails
obtained for the traditional Boltzmann equation \cite{ep,van}. The
enhancement in the likelihood of finding energetic particles is
due to the effective reduction in their collision rate.

Typically, in kinetic theory of molecular gases, the velocity
distributions have sharp tails such that all moments of the
distribution are finite\footnote{Algebraic high-energy tails may
also characterize {\em nonequilibrium} states of elastic gases;
for uniform shear flows of two-dimensional Maxwell molecules, this
has been recently shown by Acedo, Santos, and Bobylev
\cite{bobyl}.}. Sonine expansions, where Maxwellian distribution
are systematically modified by polynomials of increasing orders,
are widely used to analyze the Boltzmann equation. The Maxwell
model can be conveniently utilized to examine the applicability of
this approach to situations where the velocity distributions have
overpopulated high energy tails.

The scaling function does not characterize all features of the
asymptotic time dependent behavior. Higher than third moments of
the scaling function diverge, and since moments must be finite at
all times, a direct calculation is necessary. Moments of the
velocity distribution, $M_n(t)=\int dv\, v^n P(v,t)$, obey a
closed set of equations
\begin{equation}
\label{mom-eq-1d} {d\over d\tau}\,M_n+a_nM_n =\sum_{m=2}^{n-2}
{n\choose m} p^m(1-p)^{n-m}M_mM_{n-m},
\end{equation}
with $a_n(p)=1-p^n-(1-p)^n$.  These equations are solved
recursively using $M_0=1$ and $M_1=0$. Assuming that all moments
are initially finite we find that to leading order, the moments
asymptotically decay as
\begin{equation}
\label{mom-1d} M_n\sim e^{-a_n\tau}\sim t^{-2a_n/a_2}\,.
\end{equation}
Indeed, the sum term in (\ref{mom-eq-1d}) is asymptotically
negligible because the coefficients satisfy the inequality
$a_n<a_m+a_{n-m}$ for all $1<m<n-1$ \cite{bk}. Thus, the
temperature does not characterize higher order moments. Instead,
$M_n\sim M_2^{\alpha_n}$ with a nonlinear multiscaling spectrum
$\alpha_n=a_n/a_2$, rather than the linear spectrum $\alpha_n=n/2$
that characterizes ordinary scaling behavior. The indices
$\alpha_n$ increase monotonically with $p$, so the stronger the
dissipation, the more pronounced the multiscaling asymptotic
behavior.

The scaling behavior characterizes only velocities of the scale of
the typical velocity. For sufficiently large velocities, far
outside the scaling region, the gain term in the Boltzmann
equation is negligible, and $({\partial\over \partial
\tau}+1)P(v,\tau)=0$. Consequently, a generic exponential decay,
$P(v,\tau)\sim P_0(v)\exp(-\tau)$ characterizes such large
velocities. Similarly, sufficiently large moments decay according
to $M_n\sim e^{-\tau}\sim t^{-1/[p(1-p)]}$ for $n\to\infty$.

In several studies, inelasticity is treated as a small
perturbation \cite{plmv,bbrtv,gr}.  To first order in $p$,
Eq.~(\ref{four-eq-1d}) reads $\left({\partial\over \partial
\tau}+pk{\partial\over \partial k}\right) F(k,\tau)=0$. The
solution to this equation, $F(k,\tau)=F_0(ke^{-p\tau})$, remembers
the initial conditions forever, in contradiction with the exact
asymptotic behavior (\ref{four-scl-1d}).  This example raises
questions concerning the validity of perturbation analysis in the
vicinity of $p=0$. The $p\to 0$ limit is singular and indeed, in
Eq.~({\ref{four-scl-eq-1d}) the small parameter $p$ multiplies the
highest derivative.

The Fourier transform can be expressed as a series expansion. The
linear term can be eliminated from Eq.~(\ref{four-eq-1d}) by
making the transformation $F\to e^{\tau}F$ and $\tau\to
1-e^{-\tau}$.  Then a formal Taylor expansion solution can be
found:
\begin{equation}
\label{four-exp-1d} F(k,\tau)=e^{-\tau}\sum_{n=0}^{\infty}
{(1-e^{-\tau})^n\over n!}\,F_n(k).
\end{equation}
The expansion functions $F_n(k)$ are obtained recursively:
\hbox{$F_0(k)\equiv F(k,\tau=0)$}, and generally
$F_{n+1}(k)=\sum_{m=0}^n {n\choose m} F_m(k-pk)F_{n-m}(pk)$. For
instance, when the initial distribution is Maxwellian, the
expansion functions, and hence the velocity distribution itself
consist of sums of Maxwellians. Similarly, starting from a
stretched exponential, all expansion functions consist of sums of
stretched exponentials. In general, the expansion functions are
products of $F_0$'s with stretched arguments. This implies that
starting from a compact initial distribution $P(v,0)$, the
velocity distribution $P(v,t)$ develops a set of singularities.
For instance, a distribution with support in $[-v_0, v_0]$ becomes
non-analytic at an infinite set of points $v_{l,m}=\pm p^l(1-p)^m
v_0$.

Thus far, we characterized velocity statistics at a specific point
in time. The autocorrelation function
\begin{equation}
\label{at-def-1d} A(t',t)=\overline{v(t')\,v(t)}
\end{equation}
with the overline denoting an average over all particles,
quantifies (two-point) temporal correlations in the velocity of a
tagged particle. We have
\begin{eqnarray}
\label{at-exp} {d\over d\tau}\,A(\tau',\tau)
&=&\overline{v(\tau')\,[d v(\tau)/d\tau]}
=-(1-p)\,\overline{v(\tau')\,[v(\tau)-u(\tau)]}\nonumber\\
&=&-(1-p)\,A(\tau',\tau).
\end{eqnarray}
Therefore, the autocorrelation decays exponentially with the
collision number. In terms of the original time variable:
\begin{equation}
\label{at-1d}
A(t',t)=A_0\left(1+t'/t_0\right)^{1/p-2}\left(1+t/t_0\right)^{-1/p},
\end{equation}
with $A_0=T_0$. In particular, $A(t)\equiv A(0,t)\propto
t^{-1/p}$, so memory of the initial conditions decays
algebraically with time.  The autocorrelation function decays
faster than the temperature, $A(t)\leq T(t)$ (the two functions
coincide in the completely inelastic case, $p=1/2$). Generally,
the autocorrelation is a function of the waiting time $t'$ and
the observation time $t$, and not simply of their difference,
$t-t'$.  This history dependence (``aging'') merely reflects the
fact that the collision rate keeps changing with time.

The spread in the position of a tagged particle,
$\Delta^2(t)\equiv\langle |x(t)-x(0)|^2\rangle$, is obtained from
the autocorrelation function using $\Delta^2=2\int_0^t
dt'\int_0^{t'} dt''A(t'',t')$. Substituting the autocorrelation
function (\ref{at-1d}), we find that asymptotically, the spread
grows logarithmically with time
\begin{equation}
\label{spread} \Delta\propto \sqrt{\ln t}.
\end{equation}
This behavior reflects the $t^{-1}$ decay of the overall velocity
scale \cite{bp,kh}.

\subsection{The forced case}

In experimental situations, granular ensembles are constantly supplied with
energy, typically through the boundaries, to counter the dissipation
occurring during collisions\footnote{When the mean-free path is comparable
  with the system size, the boundary effectively plays the role of a thermal
  heat bath.}.  Theoretically, it is convenient to consider uniformly heated
systems, as discussed by Williams and MacKintosh \cite{mac} in the case of
one-dimensional inelastic hard rods and then extended to higher dimensions
(see \cite{van,plmv,bbrtv} and references therein).  We therefore study a
system where in addition to changes due to collisions, velocities also change
due to an external forcing: \hbox{${dv_j\over dt}|_{\rm heat}=\xi_j$}. We
assume standard uncorrelated white noise: $\langle \xi_j\rangle=0$ and
\hbox{$\langle\xi_i(t)\xi_j(t')\rangle=2\overline{D}\delta_{ij}\delta(t-t')$}.

Such white noise forcing amounts to diffusion with a `diffusion'
coefficient $\overline{D}$ in velocity space. Therefore, the
Boltzmann equation (\ref{be-1d}) is modified by a diffusion term
\begin{eqnarray}
\label{be-std-1d} \left({\partial \over\partial t}
-\overline{D}\,{\partial^2 \over\partial v^2}\right)P(v,t)
&=&K\int du_1\,  P(u_1,t)\int du_2\, P(u_2,t) \\
&&\times\big\{\delta\left[v-u_1+(1-p)g\right]-\delta(v-u_2)\big\}\nonumber
\end{eqnarray}
The temperature changes according to $dT/dt+\lambda
T^{3/2}=2\overline{D}$, so the steady state temperature is
$T_\infty=(2\overline{D}/\lambda)^{2/3}$. The relaxation toward
the steady state is exponential, $|T-T_\infty|\sim \exp(-{\rm
const.}\times t)$.

At the steady state, the Fourier transform $F_\infty(k)\equiv
F(k,\infty)$ satisfies
\begin{equation}
\label{std-eq-1d}
(1+Dk^2)F_\infty(k)=F_\infty(k-pk)\,F_\infty(pk).
\end{equation}
with $D=\overline{D}/\sqrt{T_\infty}$. Conservation of the total
number of particles and the total momentum impose $F_\infty(0)=1$
and $F_\infty'(0)=0$, respectively. The solution is found
recursively to give the following infinite product \cite{bk}
\begin{equation}
\label{std-sol-1d} F_\infty(k)=\prod_{l=0}^\infty\prod_{m=0}^l
\left[1+p^{2m}(1-p)^{2(l-m)}Dk^2\right]^{-{l\choose m}}.
\end{equation}
Thus in one dimension the Fourier transform in determined
analytically in the steady state.  To extract the high-energy tail
from (\ref{std-sol-1d}) we note that $F_\infty(k)$ has an infinite
series of poles located at \hbox{$\pm
i\left[p^{2m}(1-p)^{2(l-m)}\, D\right]^{-1/2}$}.  The simple poles
at $\pm i/\sqrt{D}$ closest to the origin imply an exponential
decay of the velocity distribution \cite{erb,droz},
\begin{equation}
\label{dis-std-tail-1d} P_\infty(v)\simeq {A(p)\over v_*}
\,\,e^{-|v|/v_*}, \qquad {\rm with}\qquad v_*=\sqrt{D},
\end{equation}
when $|v|\to\infty$. A re-summation similar to that used in
\cite{bk} can be performed to evaluate the residue to this pole,
and in turn, the prefactor
\begin{equation}
\label{A} A(p)={1\over 2}\,\exp\left(\sum_{n=1}^\infty {1\over
n}\,{1-a_{2n}(p)\over a_{2n}(p)}\right).
\end{equation}
This prefactor rapidly diverges, $A(p)\propto \sqrt{p}\,
\exp\left[\pi^2/(12\,p)\right]$, as $p\to 0$.  We have obtained
only the leading asymptotic behavior. It will be interesting to
determine for what range of velocities this behavior actually sets
in.

The leading high-energy behavior can also be derived by using a
useful heuristic argument \cite{ep,van,erb}. For sufficiently
large velocities, the gain term in the collision integral in
Eq.~(\ref{be-std-1d}) is negligible.  The resulting equation for
the steady state distribution
\begin{equation}
\label{Pinfty} D\,\frac{d^2 }{d v^2}P_\infty(v)=-P_\infty(v)
\end{equation}
yields the exponential high-energy tail (\ref{dis-std-tail-1d}).
This argument applies to arbitrary collision rates. For example,
if $K\propto v^\delta$, the right-hand side in (\ref{Pinfty})
becomes $-v^\delta P_\infty$ implying that $P_\infty(v)\propto
\exp(-|v|^\gamma)$ with $\gamma=1+\delta/2$.  For hard spheres
($\delta=1$) one finds $\gamma=3/2$ \cite{van}, and curiously, the
Gaussian tail arises only for the so-called very hard spheres
($\delta=2$) \cite{e}.

Finally, we notice that steady state properties in the heated case
are intimately related with the relaxation properties in the
cooling case. This can be seen via the cumulant expansion
\begin{equation}
\label{cumulant} F_\infty(k)=\exp\left[\sum_{n=1}^{\infty}
\psi_n(-Dk^2)^n\right]\,.
\end{equation}
Replacing the term $(1+Dk^2)$ with $\exp[-\sum_{n=1}
n^{-1}(-Dk^2)^n]$, and substituting the cumulant expansion into
Eq.~(\ref{std-eq-1d}) yields $\psi_n=[na_{2n}(p)]^{-1}$. The
cumulants $\kappa_n$ are defined via $\ln
F_\infty(k)=\sum_{n=1}\kappa_n(ik)^n/n!$. Therefore, the
steady-state cumulants are directly related to the relaxation
coefficients (\ref{mom-1d}), $\kappa_{2n}={(2n-1)!\over
n}{D^n\over a_{2n}}$ (the odd cumulants vanish).

\section{Uniform Gases: Arbitrary Dimension}

\subsection{The freely cooling case}

In general dimension, the colliding particles exchange momentum
only along the impact direction. The post-collision velocities
${\bf v}_{1,2}$ are given by a linear combination of the
pre-collision velocities ${\bf u}_{1,2}$,
\begin{equation}
\label{rule-d} {\bf v}_{1,2}={\bf u}_{1,2}\mp(1-p)\,({\bf g}\cdot
{\bf n})\,{\bf n}.
\end{equation}
Here ${\bf g}={\bf u}_1-{\bf u}_2$ is the relative velocity and
${\bf n}$ the unit vector connecting the particles' centers. The
normal component of the relative velocity is reduced by the
restitution coefficient $r=1-2p$, and the energy dissipation is
given by $\Delta E = -p(1-p)({\bf g}\cdot{\bf n})^2$.

In random collision processes, both the impact direction and the
identity of the colliding particles are chosen randomly.  In such
a process, the Boltzmann equation
\begin{eqnarray}
\label{be-d} {\partial P({\bf v},t)\over \partial t}=&&K \int
d{\bf
n} \int d{\bf u}_1\, P({\bf u}_1,t) \int d{\bf u}_2\,P({\bf u}_2,t)\\
&&\times \big\{\delta\left[{\bf v}-{\bf u}_1 +(1-p)({\bf g}\cdot
{\bf n}){\bf n}\right]-\delta({\bf v}-{\bf u}_1)\big\}\nonumber
\end{eqnarray}
{\em exactly} describes the evolution of the velocity distribution
function $P({\bf v},t)$.  The overall collision rate is chosen to
represent the typical relative velocity, $K=\sqrt{T}$, with the
granular temperature now being the average velocity fluctuation
per degree of freedom, $T={1\over d}\int d{\bf v}\, v^2 P({\bf
v},t)$ with $v\equiv|{\bf v}|$. The evolution equation involves
integration over all impact directions, and this angular
integration should be normalized, $\int d{\bf n}=1$.  We tacitly
ignored the restriction ${\bf g}\cdot {\bf n}>0$ on the angular
integration range in Eq.~(\ref{be-d}) since the integrand obeys
the reflection symmetry ${\bf n}\to -{\bf n}$.

Several temporal characteristics such as the temperature and the
autocorrelation behave as in the one-dimensional case.  For
example, the temperature satisfies Eq.~(\ref{temp-eq-1d}) with
prefactor $\lambda=2p(1-p)/d$ reduced\footnote{The reduction in
rate by a factor $d$ is intuitive because of the $d$ independent
directions only the impact direction is relevant in collisions.
Mathematically, the prefactor $\lambda=2p(1-p)\int d{\bf n}\,
n_1^2$ is computed by using the identity $n_1^2+\ldots+n_d^2=1$
that yields $\int d{\bf n}\, n_1^2=1/d$.} by a factor $d$. The
temperature therefore decays according to Haff's law
(\ref{temp-1d}), with the time scale
$t_0=d/\left[p(1-p)\sqrt{T_0}\right]$ set by the initial
temperature. Similarly, the decay rate of the autocorrelation
function is merely reduced by the spatial dimension,
$\frac{d}{d\tau}A(\tau',\tau)=-\frac{1-p}{d}\,A(\tau',\tau)$.
Consequently, the nonuniversal decay (\ref{at-1d}) and the
logarithmic spread (\ref{spread}) hold in general.

The Fourier transform, $F({\bf k},t)=\int d{\bf v}\,e^{i{\bf
k}\cdot {\bf v}}\,P({\bf v},t)$, satisfies
\begin{equation}
\label{four-eq-d} {\partial \over \partial \tau}\,F({\bf
k},\tau)+F({\bf k},\tau)= \int d{\bf n}\,F({\bf k}-{\bf
p},\tau)\,\,F({\bf p},\tau)
\end{equation}
with ${\bf p}=(1-p)({\bf k}\cdot {\bf n})\,{\bf n}$ reflecting the
momentum transfer occurring during collisions.  This equation was
obtained by multiplying Eq.~(\ref{be-d}) by $e^{i{\bf k}\cdot {\bf
v}}$ and integrating over the velocities.  The power of the
Fourier transform is even more remarkable in higher
dimensions\footnote{For elastic Maxwell molecules, the
  Fourier transform was first used in unpublished theses
  by Krupp \cite{krupp}, and then rediscovered and successfully
  utilized by Bobylev (see \cite{e,bob} for a review).}  as it reduces the
$(3d-1)-$fold integral in Eq.~(\ref{be-d}) to the $(d-1)-$fold
integral in Eq.~(\ref{four-eq-d}).

Hereinafter, we consider only isotropic velocity distributions.
The Fourier transform depends only on $k\equiv |{\bf k}|$, so we
write $F({\bf
  k},\tau)=F(y,\tau)$ with $y=k^2$. To perform the angular integration, we
employ spherical coordinates with the polar axis parallel to ${\bf
k}$, so that $\hat{\bf k}\cdot {\bf n}=\cos \theta$. The
$\theta$-dependent factor of the angular integration measure
$d{\bf n}$ is proportional to $(\sin \theta)^{d-2} d\theta$.
Denoting angular integration with brackets, $\langle f
\rangle\equiv \int d{\bf n}f $, and using $\mu=\cos^2\theta$ gives
\begin{equation}
\label{av-def} \langle f\rangle =\int_0^1 d\mu\,
 {\mu^{-{1\over2}} (1-\mu)^{d-3\over 2}\over
B\big({1\over 2},{d-1\over 2}\big)}\, f(\mu),
\end{equation}
where $B(a,b)$ is the beta function. This integration is properly
normalized, $\langle 1 \rangle =1$. The governing equation
(\ref{four-eq-d}) for the Fourier transform can now be rewritten
in the compact from
\begin{equation}
\label{fkt1} {\partial \over \partial \tau}\,F(y,\tau)+F(y,\tau)=
\big\langle \,F(\xi y,\tau)\,F(\eta y,\tau)\,\big\rangle,
\end{equation}
with the shorthand notations $\xi=1-(1-p^2)\mu$ and
$\eta=(1-p)^2\mu$. Unlike the one-dimensional case, explicit
solutions of this nonlinear and nonlocal rate equation are
cumbersome and practically useless. Nevertheless, most of the
physically relevant features of the velocity distributions
including the large velocity statistics and the time dependent
behavior of the moments can be obtained analytically.

We seek a scaling solution: $P({\bf v},t)=T^{-d/2}{\cal P}(w)$
with $w=vT^{-1/2}$, or equivalently $F(y,\tau)=\Phi(x)$ with
$x=yT$. The scaling function $\Phi(x)$ satisfies
\begin{equation}
\label{four-scl-eq-d} -\lambda x\,\Phi'(x)+\Phi(x) =\big\langle
\Phi(\xi x)\,\Phi(\eta x)\big\rangle
\end{equation}
and the boundary condition $\Phi(x)=1-{1\over 2}\,x+\cdots$ as
$x\to 0$.  In the elastic case, the velocity distribution is
purely Maxwellian, $\Phi(x)=e^{-x/2}$. Indeed, $\lambda=0$ and
$\xi+\eta=1$ in this case. A stochastic process of elastic
collisions effectively randomizes the velocities and leads to a
thermal distribution \cite{max}. In practice, this collision
algorithm is used in Molecular dynamics simulations to thermalize
velocities.

{}From the one-dimensional case, we anticipate that the velocity
distribution has an algebraic large velocity tail. Generally, the
large velocity behavior can be determined from the small wave
number behavior of the Fourier transform. For example, the
small-$x$ expansion of the one-dimensional solution
(\ref{dis-scl-1d}) contains both regular and singular terms:
\hbox{$\Phi(x)=1-{1\over 2}\,x+{1\over 3}\,x^{3/2}+\cdots$}, and
the dominant singular $x^{3/2}$ term reflects the $w^{-4}$ tail of
${\cal P}(w)$.  In general, if ${\cal P}(w)$ has an algebraic
tail,
\begin{equation}
\label{dis-tail} {\cal P}(w)\sim w^{-\sigma}\qquad {\rm as}\quad
w\to\infty,
\end{equation}
then \hbox{$\Phi(x)\propto \int_0^\infty dw\, w^{d-1}{\cal P}(w)\,
  e^{iw\sqrt{x}}$} contains, apart from regular terms, the following
  dominant singular term: $\Phi_{\rm sing}(x)\sim x^{(\sigma-d)/2}$ as
  $x\to 0$.  The exponent $\sigma$ can be now obtained by inserting
  $\Phi(x)=\Phi_{\rm reg}(x)+\Phi_{\rm sing}(x)$ into
  Eq.~(\ref{four-scl-eq-d}) and balancing the dominant singular
  terms. We find that $\sigma$ is a root of the integral 
equation\footnote{This result was derived independently in \cite{kb,eb1}.}
\begin{equation}
\label{sigma-d} 1-\lambda\,{\sigma-d\over 2} =\big\langle
\xi^{(\sigma-d)/2}+\eta^{(\sigma-d)/2}\big\rangle.
\end{equation}
This equation can be recast as the eigenvalue problem
$\lambda_{\nu}=\nu\lambda_1$ using $\lambda_{\nu}=\langle
1-\xi^\nu-\eta^\nu\rangle$ and $\nu=(\sigma-d)/2$. It can also be
expressed using the hypergeometric function ${}_2F_1(a,b;c;z)$
\cite{aa} and the Euler's gamma function:
\begin{eqnarray*}
\label{sigma-trans} 1-p(1-p){\sigma-d\over d}= {}_2F_1
\left[{d-\sigma\over 2},{1\over2};{d\over 2}; 1-p^2\right]
+(1-p)^{\sigma-d}\, {\Gamma\left({\sigma-d+1\over
2}\right)\Gamma\left({d\over 2}\right)\over
\Gamma\left({\sigma\over 2}\right)\,\Gamma\left({1\over
2}\right)}.
\end{eqnarray*}
Clearly, the exponent $\sigma\equiv \sigma(d,p)$ depends in a
nontrivial fashion on the spatial dimension $d$ as well as the
collision parameter $p$.

\begin{figure}[t]
\centerline{\includegraphics[width=7.6cm]{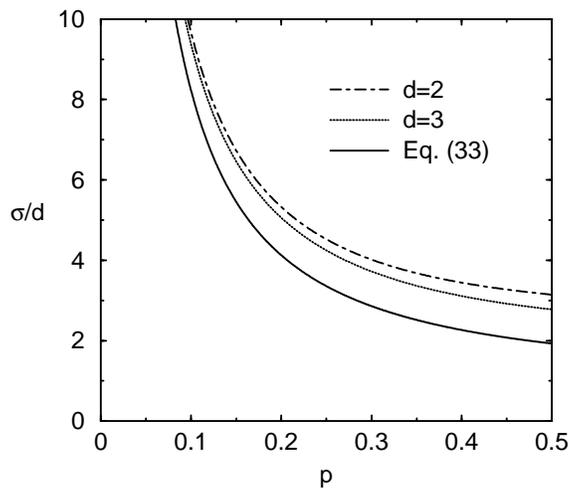}} \caption{The
exponent $\sigma$ versus the collision parameter $p$ for $d=2$,
and $3$. For comparison with the leading large-dimension behavior
(\ref{sigma-lim2}), $\sigma$ is rescaled by $d$.}
\end{figure}

There are two limiting cases where the velocity distribution
approaches a Maxwellian, thereby implying a divergence of the
exponent $\sigma$. In the quasi-elastic limit ($p\to 0$) we have
$\sigma\to d/p$. Even a minute degree of dissipation strongly
changes the character of the system, and hence, energy dissipation
is a singular perturbation \cite{bcdr,plmv}. As the dimension
increases, the relative weight of the impact direction diminishes,
and so does the role of dissipation. For large dimensions, the
integral $\langle \eta^{(\sigma-d)/2}\rangle$ in
Eq.~(\ref{sigma-d}) vanishes exponentially and asymptotic analysis
of the remaining integral shows that the exponent grows linearly
with the dimension:
\begin{equation}
\label{sigma-lim2} \sigma\to d \, {1+{3\over 2}p-p^3-p^{1/2}
\left(1+{5\over 4}p\right)^{1/2}\over p(1-p^2)}
\end{equation}
when $d\to\infty$ \cite{kb}. Equation (\ref{sigma-lim2}) provides
a decent approximation even at moderate dimensions (Fig.~1).
Overall, the exponent $\sigma(d,p)$ increases monotonically with
increasing $d$, and additionally, it increases monotonically with
decreasing $p$. Both features are intuitive as they mirror the
monotonic dependence of the energy dissipation rate
$\lambda=2p(1-p)/d$ on $d$ and $p$. Remarkably, the exponent is
very large. The completely inelastic case provides a lower bound
for the exponent, $\sigma(d,p)\ge \sigma(d,1/2)$ with
$\sigma(d,1/2)=6.28753$, $8.32937$, for $d=2$, $3$, respectively.
For typical granular particles, $\sigma(d=3,p=0.1)\cong 30$, and
such algebraic decays are impossible to measure in practice.

The algebraic tail of the velocity distribution implies that
moments of the scaling function $\Phi(x)$ with sufficiently large
indices diverge. In the scaling regime, moments of the velocity
distribution can be calculated by expanding the Fourier transform
in powers of $x$,
\begin{equation}
\label{four-exp-d} \Phi(x)\cong\sum_{n=0}^{[\nu]} \phi_n (-x)^n.
\end{equation}
The order of terms in this expansion must be smaller than the
order of the singular term, $x^\nu$. The coefficients $\phi_n$,
needed for calculating transport coefficients \cite{santos}, yield
the leading asymptotic behavior of the velocity moments,
$M_k(t)=\int d{\bf v}\, v^k P({\bf v},t)$, via the relation
$(2n)!\,T^n\phi_n\simeq \langle \mu^n\rangle M_{2n}$. Inserting
the moment expansion into (\ref{four-scl-eq-d}) yields a closed
hierarchy of equations
\begin{eqnarray}
\label{four-rec-d} (\lambda_n-n\lambda_1)\phi_n=\sum_{m=1}^{n-1}
\lambda_{m,n-m} \phi_m\phi_{n-m},
\end{eqnarray}
with $\lambda_n=\langle 1-\xi^n-\eta^n\rangle$ and
$\lambda_{m,l}=\langle\xi^m\eta^l\rangle$. Calculation of these
coefficients requires the following integrals
\begin{eqnarray*}
\langle \mu^n\rangle=
{\Gamma\left({d\over2}\right)\Gamma\left(n+{1\over 2}\right)\over
\Gamma\left({1\over2}\right)\Gamma\left(n+{d\over 2}\right)}=
{1\over d}{3\over d+2}\cdots{2(n-1)+1\over 2(n-1)+d}.
\end{eqnarray*}
Of course, $\phi_0=1$ and $\phi_1=1/2$; further coefficients can
be determined recursively from (\ref{four-rec-d}), e.g.,
\begin{equation}
\label{phi2} \phi_2={1\over 8}\,{1-3\,{1-p^2\over d+2}\over
1-3\,{1+p^2\over d+2}}.
\end{equation}
This coefficient is finite only when $\lambda_2>2\lambda_1$ or
$d<d_2=1+3p^2$. Generally $\phi_n$ is finite only when the left
hand side of Eq.~(\ref{four-rec-d}) is positive,
$\lambda_n>n\lambda_1$, or equivalently, when the dimension is
sufficiently small, $d<d_n$. The crossover dimensions $d_n$ are
determined from $\lambda_n=n\lambda_1$.

For $d>d_n$, moments with index smaller than $2n$ are
characterized by the temperature, while higher moments exhibit
multiscaling asymptotic behavior.  The time evolution of the
moments can be studied using the expansion
\begin{equation}
\label{mom-exp-d} F(y,\tau)=\sum_{n=0}^\infty f_{n}(\tau)\,(-y)^n.
\end{equation}
The actual moments are related to the $f_n$ coefficients via
$(2n)!f_{n}=\langle \mu^n\rangle M_{2n}$.  Substituting the
expansion (\ref{mom-exp-d}) into (\ref{fkt1}) yields the evolution
equations
\begin{equation}
\label{mom-rec-d} {d\over d\tau}
f_{n}+\lambda_{n}f_{n}=\sum_{m=1}^{n-1}\lambda_{m,n-m}f_{m}f_{n-m}.
\end{equation}
We have $f_1\propto e^{-\lambda_1\tau}$, this is just the Haff's
law. {}From $df_2/d\tau+\lambda_2f_2=\lambda_{1,1}f_1^2$ we see
that $f_2$ is a linear combination of two exponentials,
$e^{-\lambda_2\tau}$ and $e^{-2\lambda_1\tau}$. The two decay
coefficients are equal $\lambda_2=2\lambda_1$ at the crossover
dimension $d=d_2$. As expected, when $d<d_2$ the fourth moment is
dictated by the second moment, $f_2\propto f_1^2$. Otherwise,
$f_2\propto\exp(-\lambda_2\tau)$. In general,
\begin{equation}
\label{mom-eq-d} M_{2n}\propto \cases{\exp(-n\lambda_1\tau)
&$d<d_n$;\cr
                     \exp(-\lambda_n\tau)  &$d>d_n$.}
\end{equation}
Fixing the dimension and the collision parameter, moments of
sufficiently high order exhibit multiscaling asymptotic behavior.
In practice, the exponent $\sigma$ is typically large, and
multiscaling occurs only for very high order moments.

\subsection{The forced case}

We consider white noise forcing as in the one-dimensional case.
The steady state Fourier transform, $F_\infty(k)\equiv
F_\infty(k^2)$, satisfies
\begin{equation}
\label{four-std-eq-d} (1+Dk^2)F_\infty(k^2)=\big\langle
F_\infty(\xi k^2)\, F_\infty(\eta k^2)\big\rangle.
\end{equation}
This equation is solved recursively by employing the cumulant
expansion (\ref{cumulant}). Writing $1+Dk^2=\exp\left[\sum_{n\geq
1}(-Dk^2)^n/n\right]$, we recast Eq.~(\ref{four-std-eq-d}) into
\begin{eqnarray}
\label{cumeqg} 1=\Bigg\langle\exp\left[-\sum_{n=1}^{\infty}
\left(\widetilde \psi_n-n^{-1}\right)(-Dk^2)^n\right]\Bigg\rangle,
\end{eqnarray}
with the auxiliary variables
$\widetilde\psi_n=\psi_n(1-\xi^n-\eta^n)$.  The coefficients
$\psi_n$ are obtained by evaluating recursively the angular
integrals of the auxiliary variables, $\langle
\widetilde\psi_n\rangle$, and then using the identities
$\psi_n=\langle \widetilde\psi_n\rangle/\lambda_n$. The few first
coefficients can be determined explicitly, e.g.,
$\psi_1=1/\lambda$ and
\begin{equation}
\label{psi2} \psi_2={3\,d^2\over 4(d+2)(1-p^2)-12(1-p)^2(1+p^2)}.
\end{equation}
The nonvanishing second cumulant shows the steady state
distribution is not Maxwellian. Moreover, the poles of the Fourier
transform located at $k=\pm i/\sqrt{D}$ indicate that the large
velocity tail of the distribution is exponential as in the
one-dimensional case (\ref{dis-std-tail-1d}). Exponential decay is
also suggested by the heuristic argument detailed above.

\subsection{Velocity correlations}

Maxwellian velocity distributions were originally obtained for
random elastic collision processes (see Ref.~\cite{classical},
p.~36). Maxwell's seminal derivation involves two basic
assumptions: (1) The velocity distribution is isotropic, and (2)
Correlations between the velocity components are absent.  In
inelastic gases, the velocity distributions are non-Maxwellian --
therefore, there must be correlations between the velocity
components\footnote{Inelastic collisions discriminate the impact
direction, thereby generating correlations among the velocity
components.}.  The quantity
\begin{equation}
\label{corr-def} U={\langle v_x^2v_y^2\rangle-\langle
v_x^2\rangle\langle v_y^2\rangle \over \langle v_x^2\rangle\langle
v_y^2\rangle}
\end{equation}
provides a natural correlation measure.  A non-vanishing $U$
indicates that velocity correlations do exist, and the larger $U$
the larger the correlation.

\begin{figure}[t]
\centerline{\includegraphics[width=7.6cm]{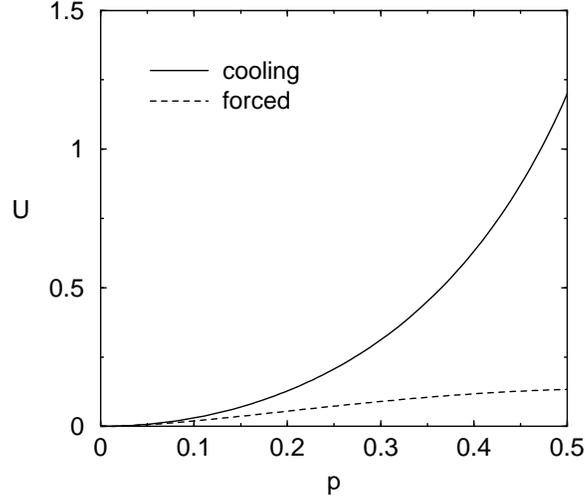}} \caption{The
correlation measure $U$ for spatial dimension $d=3$. Shown are the
freely cooling case and the forced case.}
\end{figure}

To compute $U$, we apply the identities: $\langle
v_x^2\rangle={\partial^2\over\partial k_x^2}\,F\Big|_{{\bf k}={\bf
0}}$ and $\langle v_x^2v_y^2\rangle={\partial^2\over\partial
k_x^2} {\partial^2\over\partial k_y^2}\,F\Big|_{{\bf k}={\bf 0}}$.
In the freely cooling case, $U=8\phi_2-1$; in the forced case,
$U=\psi_2/\psi_1^2$. Substituting the corresponding coefficients
yields
\begin{subeqnarray}
U_{\rm cooling}&=&\frac{6p^2}{d-(1+3p^2)}, \label{corra}\\
U_{\rm forced}&=&\frac{6p^2(1-p)}{(d+2)(1+p)-3(1-p)(1+p^2)}.
\label{corrb}
\end{subeqnarray}
Generally, correlations increase monotonically with $p$.
Correlations are much weaker in the presence of an energy source
because the nature of the driving is random (Fig.~2).  The
perfectly inelastic case provides an upper bound. For example,
when $d=3$ and $p=1/2$ we have $U_{\rm cooling}=6/5$ and $U_{\rm
forced}=2/15$. Correlations decay as $U\propto p^2$ and $U\propto
d^{-1}$ in the respective limiting cases of $p\to 0$ and
$d\to\infty$, where Maxwellian distributions are recovered.

\section{Impurities}

The dynamics of an impurity immersed in a uniform granular fluid
often defies intuition. The impurity does not generally behave as
a tracer particle, and instead it may move either faster or slower
than the fluid. Despite extensive studies, many theoretical and
experimental questions regarding the dynamics of impurities remain
open \cite{rsps,jmp,kjn,dj,hql,mlnj,dufty}.

Impurities represent the simplest form of polydispersity, an
important characteristic of granular media \cite{jcw,csc}.
Theoretically, the impurity problem is a natural first step in the
study of mixtures as it involves fewer collision parameters. The
fluid background is not affected by the presence of the impurity,
and the impurity can be seen as ``enslaved'' to the fluid
background.

We study dynamics of a single impurity particle in a uniform
background of identical inelastic particles. We set the fluid
particle mass to unity and the impurity mass to $m$.  Collisions
between two fluid particles are characterized by the collision
parameter $p$ as in Eq.~(\ref{rule-d}), while collisions between
the impurity and any fluid particle are characterized by the
collision parameter $q$.  When an impurity particle of velocity
${\bf u}_1$ collides with a fluid particle of velocity ${\bf u}_2$
its post-collision velocity ${\bf v}_1$ is given by
\begin{equation}
\label{rule-imp} {\bf v}_{1}={\bf u}_{1}-(1-q)\,({\bf g}\cdot {\bf
n})\,{\bf n}.
\end{equation}
Two restitution coefficients, $r_p\equiv r=1-2p$ and
$r_q=m-(m+1)q$, characterize fluid-fluid and impurity-fluid
collisions, respectively. The restitution coefficients obey $0\leq
r\leq 1$, and the collision parameters accordingly satisfy $0\leq
p\leq 1/2$ and \hbox{${m-1\over m+1}\leq q\leq {m\over m+1}$}. The
energy dissipated in an impurity-fluid collision is $\Delta
E=-m(1-q)[2-(m+1)(1-q)]({\bf g}\cdot{\bf n})^2$.

Let $Q({\bf v},t)$ be the normalized velocity distribution of the
impurity. In a random collision process, the impurity velocity
distribution evolves according to the {\it linear}
Lorentz-Boltzmann equation
\begin{eqnarray}
\label{be-imp} {\partial Q({\bf v},t)\over \partial t} =&K_q&\int
d{\bf n} \int d{\bf u}_1\,Q({\bf u}_1,t)
\int d{\bf u}_2\,P({\bf u}_2,t) \nonumber \\
&\times& \left\{\delta\Big[{\bf v}-{\bf u}_1+(1-q)({\bf g}\cdot
{\bf n}){\bf n}\Big] -\delta({\bf v}-{\bf u}_1)\right\},
\end{eqnarray}
while the fluid distribution obeys (\ref{be-d}). Two rates,
$K_p\equiv K$ and $K_q$, characterize fluid-fluid and
fluid-impurity collisions, respectively. These rates can be eliminated from
the respective equations (\ref{be-d}) and (\ref{be-imp}) by
introducing the collision counters, $\tau_p\equiv \tau=\int_0^t
dt'K_p(t')$, and $\tau_q=\int_0^t dt' K_q(t')$. The
Lorentz-Boltzmann equation is further simplified by using the
Fourier transform, $G({\bf k},t)=\int d{\bf v}\,e^{i{\bf k}\cdot
{\bf v}}\,Q({\bf
  v},t)$
\begin{eqnarray}
\label{four-eq-imp} {\partial \over \partial \tau_q}\,G({\bf
k})+G({\bf k}) = \int d{\bf n}\,G({\bf k}-{\bf q})\,F({\bf q}),
\end{eqnarray}
with ${\bf q}=(1-q)\,({\bf k}\cdot{\bf n})\,{\bf n}$. This
equation supplements the fluid equation (\ref{four-eq-d}).

We consider two versions of the Maxwell model: An idealized case
with equal collision rates, $K_p=K_q$ (model A); and a more
physical case with collision rates proportional to appropriate
average relative velocities (model B). In the forced case, one can
obtain the impurity velocity distribution using the cumulant
expansion, and generally, velocity statistics of the impurity are
similar to the background. Below, we discuss the more interesting
freely cooling case.

\subsection{Model A}

When the two collision rates are equal, all pairs of particles are
equally likely to collide with each other. For simplicity we set
the overall rate to unity, $K_p=K_q=1$; then both collision
counters equal time, $\tau_p=\tau_q=t$.

In the freely cooling case, the fluid temperature decays
exponentially, $T(t)=T_0\exp\left[-2p(1-p)t/d\right]$. The
impurity temperature, $\Theta(t)$, defined by $\Theta(t)={1\over
d}\int d{\bf v}\, v^2 Q({\bf v},t)$, is coupled to the fluid
temperature via a linear rate equation
\begin{equation}
\label{temp-eq-imp} {d \over dt}\Theta=-{1-q^2\over d}\,\Theta
+{(1-q)^2\over d}\, T.
\end{equation}
The solution to this equation is a linear combination of two
exponentials
\begin{equation}
\label{temp-imp}
\Theta(t)=\left(\Theta_0-c\,T_0\right)\,e^{-(1-q^2)t/d}
+c\,T_0\,e^{-2p(1-p)t/d}.
\end{equation}
The constant $c={(1-q)^2/[1-q^2-2p(1-p)}]$ is simply the ratio
between the impurity temperature and the fluid temperature in the
long time limit. This generalizes the elastic fluid result
$c=(1-q)/(1+q)$ \cite{p,mp1}. Generally, the impurity and the
fluid have different energies, and this lack of equipartition is
typical to granular particles \cite{wp,fm}

There are two different regimes of behavior. When $1-q^2>2p(1-p)$,
the impurity temperature is proportional to the fluid temperature
asymptotically, ${\Theta(t)\over T(t)}\to c$ as $t\to \infty$.  In
the complementary region $2p(1-p)>1-q^2$ the ratio of the fluid
temperature to the impurity temperature vanishes. Since the system
is governed by three parameters $(m,r_p,r_q)$, it is convenient to
consider the restitution coefficients as fixed and to vary the
impurity mass. From the definition of the restitution coefficients
the borderline case, $q^2=1-2p(1-p)$, defines a critical impurity
mass
\begin{equation}
\label{mstar} m_*={r_q+\sqrt{(1+r_p^2)/2}\over
1-\sqrt{(1+r_p^2)/2}}.
\end{equation}
This critical mass is always larger than unity.  Asymptotically,
the impurity to fluid temperature ratio exhibits two different
behaviors
\begin{equation}
\label{regimes} {\Theta(t)\over T(t)}\to \cases{c &$m<m_*$,\cr
       \infty   &$m\geq m_*$.\cr}
\end{equation}
We term these two regimes, the light impurity phase and the heavy
impurity phase, respectively.  In the light impurity phase, the
initial impurity temperature becomes irrelevant asymptotically,
and the impurity is governed by the fluid background. In the heavy
impurity phase, the impurity is infinitely more energetic compared
with the fluid and practically, it sees a static fluid.
Interestingly, the dependence on the dimension is secondary as
both the critical mass, $m_*$, and the temperature ratio $c$ are
independent of $d$. Below we study velocity statistics including
the velocity distribution and its moments primarily in one
dimension, where explicit solutions are possible.

\subsubsection{The light impurity phase}

In this phase, $m<m_*$, the impurity is governed by the fluid
background. We therefore seek scaling solutions of the same form
as the fluid's: $Q(v,t)= T^{-1/2}{\cal Q}\left(vT^{-1/2}\right)$
and $G(k,t)=g\left(|k|T^{1/2}\right)$.  {}From the
Lorentz-Boltzmann equation (\ref{four-eq-imp}), the latter scaling
function satisfies the {\it linear} equation
\begin{equation}
\label{four-scl-eq-imp} -p(1-p)z g'(z)+g(z)=g(q
z)\,[1+(1-q)z]\,e^{-(1-q)z}.
\end{equation}
Since the fluid scaling function is a combination of $z^ne^{-z}$
with $n=0$ and $n=1$, we try the series ansatz
\begin{equation}
\label{four-scl-exp-imp} g(z)=\sum_{n=0}^\infty g_n z^ne^{-z}
\end{equation}
for the impurity.  The first few coefficients, $g_0=g_1=1$ and
$g_2=(1-c)/2$, follow from the small-$z$ behavior $g(z)\cong
1-{1\over 2}cz^2$. The rest are obtained recursively
\begin{equation}
\label{four-scl-coef-imp} g_n={q^{n-1}(1-q)-p(1-p)\over
1-q^n-np(1-p)}\,g_{n-1}.
\end{equation}

The Fourier transform can be inverted to obtain the impurity
velocity distribution function explicitly.  The inverse Fourier
transform of $e^{-\kappa z}$ is ${1\over \pi}{\kappa\over
\kappa^2+w^2}$; the inverse transforms of $z^{n}e^{-z}$ can be
obtained using successive differentiation with respect to
$\kappa$. Therefore, the solution is a series of powers of
Lorentzians
\begin{equation}
\label{qw} {\cal Q}(w)={2\over \pi}\sum_{n=2}^{\infty}Q_n {1\over
(1+w^2)^n}
\end{equation}
The coefficients $Q_n$ are linear combinations of the coefficients
$g_k$'s with $k\leq n+1$, e.g., $Q_2=1-3g_2+3g_3$ and
$Q_3=4g_2-24g_3+60g_4$. There are special values of $q$ for which
the infinite sum terminates at a finite order. Of course, when
$q=p$, the impurity is identical to the fluid and $Q_n=0$ for all
$n>2$.  When $q^2(1-q)=p(1-p)$, one has $Q_n=0$ for all $n>3$, so
the velocity distribution includes only the first two terms.
Regardless of $q$, the first squared Lorentzian term dominates the
tail of the velocity distribution ${\cal Q}(w)\sim{\cal P}(w)\sim
w^{-4}$, as $w\to \infty$.   One can show that this behavior
extends to higher dimensions, ${\cal Q}(w)\sim{\cal P}(w)\sim
w^{-\sigma}$. Therefore, the impurity has the same algebraic
extremal velocity statistics as the fluid.

While the scaling functions underlying the impurity and the fluid
are similar, more subtle features may differ.  Moments of the
impurity velocity distribution, $L_n(t)=\int dv\, v^n\,Q(v,t)$,
obey the recursive equations
\begin{eqnarray}
\label{mom-eq-imp} {d\over dt}L_n+b_nL_n &=&\sum_{m=2}^{n-2}
{n\choose m} q^{m}(1-q)^{n-m}L_{m}M_{n-m},
\end{eqnarray}
with $b_n(q)=1-q^n$.  Asymptotically, the fluid moments decay
exponentially according to $M_{n}(t)\propto e^{-a_n(p)\,t}$. Using
this asymptotic behavior, we analyze the (even) impurity moments.
The second moment, i.e. the impurity temperature, was  shown to
behave similar to the fluid temperature when $a_2(p)<b_2(q)$.  The
fourth moment behaves similarly to the fourth moment of the fluid,
$L_4\propto M_4$, when $a_4(p)<b_4(q)$. However, in the
complementary case, the fourth moment behaves differently,
$L_n(t)\propto e^{-b_4(q)t}$.  In general, when
$a_{n}(p)<b_{n}(q)$, the $n$th impurity moment is proportional to
the $n$th fluid moment, $L_{n}\propto M_{n}$. Otherwise, when
$a_{n}(p)>b_{n}(q)$, the $n$th impurity moment is no longer
governed by the fluid and
\begin{equation}
\label{mom-imp} L_n(t)\propto e^{-(1-q^n)t}.
\end{equation}

This series of transition affecting moments of decreasing order
occurs at increasing impurity masses,
\begin{equation}
\label{masses-series} m_1>m_2>\cdots>m_{\infty}.
\end{equation}
When $m\geq m_n$, the ratio $M_{2k}/L_{2k}$ diverges
asymptotically for all $k\ge n$. The transition masses
\begin{equation}
\label{masses} m_n={r_q+{1\over
2}\left[(1-r_p)^{2n}+(1+r_p)^{2n}\right]^{1\over 2n}\over
1-{1\over 2}\left[(1-r_p)^{2n}+(1+r_p)^{2n}\right]^{1\over 2n}}
\end{equation}
are found from $q^{2n}=p^{2n}+(1-p)^{2n}$ and the definitions of
the restitution coefficients. All of the transition masses are
larger than unity, so the impurity must be heavier than the fluid
for any transition to occur.  The largest transition mass is
$m_1\equiv m_*$, and the smallest transition mass is
$m_\infty=\lim_{n\to\infty}m_n=(1+r_p+2r_q)/(1-r_p)$.  Impurities
lighter than the latter mass, $m<m_\infty$, mimic the fluid
completely.

\subsubsection{The heavy impurity phase}

In this phase, $m> m_*$, the velocities of the fluid particles are
asymptotically negligible compared with the velocity of the
impurity. Fluid particles become stationary as viewed by the
impurity, and effectively, ${\bf u}_2\equiv 0$ in the collision
rule (\ref{rule-imp}):
\begin{equation}
\label{rule-lor} {\bf v}={\bf u}-(1-q)\,({\bf u}\cdot{\bf n})\,
{\bf n}.
\end{equation}
Mathematically, this process is reminiscent of a Lorentz gas
\cite{Lor}. Physically, the two processes are different. In the
granular impurity system, a heavy particle scatters of a static
background of lighter particles, while in the Lorentz gas the
scatterers are infinitely massive.

We first consider the one-dimensional case.  Setting $u_2\equiv 0$
in the  Lorentz-Boltzmann equation (\ref{be-imp}), integration
over the fluid velocity $u_2$ is trivial, $\int d
u_2\,P(u_2,t)=1$, and integration over the impurity velocity $u_1$
gives
\begin{equation}
\label{lorentz} {\partial\over \partial t}Q(v,t)+Q(v,t)={1\over q}
Q\left({v\over q}\,,\,t\right).
\end{equation}
This equation can be solved directly by considering the stochastic
process the impurity particle experiences.  In a sequence of
collisions, the impurity velocity changes according to $v_0\to q
v_0\to q^2 v_0\to \cdots$ with $v_0$ the initial velocity.  After
$n$ collisions the impurity velocity decreases exponentially,
$v_n=q^n v_0$.  Furthermore, the collision process is random, and
therefore, the probability that the impurity undergoes exactly $n$
collisions up to time $t$ is Poissonian $t^ne^{-t}/n!$. Thus, the
velocity distribution function reads
\begin{equation}
\label{pivt} Q(v,t)=e^{-t}\sum_{n=0}^{\infty}{t^n\over n!}\,
{1\over q^n}\,Q_0\left({v\over q^n}\right),
\end{equation}
where $Q_0(v)$ is the initial velocity distribution of the
impurity. The impurity velocity distribution function is a
time-dependent combination of ``replicas'' of the initial velocity
distribution. Since the arguments are stretched, compact velocity
distributions display an infinite set of singularities, a generic
feature of the Maxwell model.

In contrast to the velocity distribution, the moments $L_n(t)$
exhibit a much simpler behavior.  Indeed, from Eq.~(\ref{lorentz})
one finds that every moment is coupled only to itself, ${d\over
dt} L_n=-(1-q^n)L_n$.  Solving this equation we recover
Eq.~(\ref{mom-imp}); in the heavy impurity phase, however, it
holds for all $n$. Therefore the moments exhibit multiscaling
asymptotic behavior.  The decay coefficients, characterizing the
$n$-th moment, depend on $n$ in a nonlinear fashion.

\subsection{Model B}

For hard sphere particles, the collision rates are proportional to
the relative velocity. The overall collision rates in the Maxwell
model represent the average relative velocity, and a natural
choice is $\sqrt{\langle ({\bf v}_1-{\bf v}_2)^2\rangle }\propto
\sqrt{(T_1+T_2)/2}$. Therefore, the rates $K_p=\sqrt{T}$ and
$K_q=\sqrt{(T+\Theta)/2}$, should be used in the Boltzmann
equation (\ref{be-d}) and the Lorentz-Boltzmann equation
(\ref{be-imp}), respectively. This modification suppresses the
heavy impurity phase, although the secondary transitions
corresponding to higher order moments remain.

The fluid and the impurity temperatures obey
\begin{eqnarray*}
\label{temp-eq-imp-b}
{d \over dt}T&=&-\sqrt{T}\left[{2p(1-p)\over d}\,T\right],\nonumber\\
{d \over dt}\,\Theta&=&\sqrt{{T+\Theta\over 2}} \left[-{1-q^2\over
d}\,\Theta +{(1-q)^2\over d}\, T\right].
\end{eqnarray*}
Consequently, the temperature ratio, $S=\Theta/T$, evolves
according to
\begin{equation}
\label{ratio-mf} {1\over \sqrt{T}}\,{d \over dt}\,S
=\sqrt{{1+S\over2}} \left[-{1-q^2\over d}\,S+{(1-q)^2\over
d}\right]+{2p(1-p)\over d}\,S.
\end{equation}
The loss term, which grows as $S^{3/2}$, eventually overtakes the
gain term that grows only linearly with $S$.  Therefore, the
asymptotic temperature ratio remains finite, $S\to c$, where $c$
is the root of the cubic equation
\begin{equation}
\label{ratio-b} \sqrt{{1+c\over 2}}\left(c-{1-q\over
1+q}\right)={2p(1-p)\over 1-q^2}\,c.
\end{equation}
Consequently, there is only one phase, the light impurity phase.
Intuitively, as the impurity becomes more energetic, it collides
more often with fluid particles. This mechanism limits the growth
rate of the temperature ratio.  As in model A, energy
equipartition does not generally occur, and the behavior is
largely independent of the spatial dimension.

Qualitatively, results obtained for the light impurity phase in
model A extend to model B.  The impurity velocity distribution
follows a scaling asymptotic behavior. The only difference is that
the collision terms are proportional to
$\beta^{-1}=\sqrt{(1+c)/2}$, and Eq.~(\ref{four-scl-eq-imp})
generalizes as follows
\begin{equation}
\label{four-scl-eq-imp-b} -\beta p(1-p)z g'(z)+g(z)=(1+\alpha
z)\,e^{-\alpha z}\,g(q z).
\end{equation}
Seeking a series solution of the form (\ref{four-scl-exp-imp})
leads to the following recursion relations for the coefficients
\begin{equation}
\label{four-scl-coef-imp-b} g_n={(1-q)q^{n-1}-\beta p(1-p)\over
1-q^n-n\beta p(1-p)}\,g_{n-1}.
\end{equation}
Again, the velocity distribution is a combination of powers of
Lorentzians as in Eq.~(\ref{qw}). Moreover, the coefficients $Q_n$
are linear combinations of the coefficients $g_n$'s as in model A.
Most importantly, the large-velocity tail is generic ${\cal
Q}(w)\sim w^{-4}$.

The fluid moments evolve according to (\ref{mom-eq-1d}) with
$\tau\equiv \tau_p$, while the impurity moments evolve according
to
\begin{eqnarray}
\label{mom-eq-imp-b} {d\over d\tau_q}L_n+b_nL_n
&=&\sum_{m=2}^{n-2} {n\choose m} q^{m}(1-q)^{n-m}M_{m}L_{m-j},
\end{eqnarray}
Note that asymptotically $\tau_q\to\tau_p/\beta$, so the fluid
moments decay according to $M_n\propto e^{-\beta a_n(p)\tau_q}$.
Hence, when $\beta a_n(p)<b_n(q)$, the impurity moments are
enslaved to the fluid moments, i.e., $L_n\propto M_n$
asymptotically. Otherwise, sufficiently large impurity moments
behave differently than the fluid moments, viz.  $M_n\propto
e^{-b_n\tau_q}$. Although the primary transition affecting the
second moment does {\em not} occur ($m_1\equiv m_*=\infty$),
secondary transitions affecting larger moments do occur at a
series of masses, as in Eq.~(\ref{masses-series}). The transition
masses $m_n$ are found by solving $\beta a_{n}(p)=b_{n}(q)$
simultaneously with Eq.~(\ref{ratio-b}). For example, for
completely inelastic collisions ($r_p=r_q=0$) one finds
$m_2=1.65$. These transitions imply that some velocity statistics
of the impurity, specifically large moments, are no longer
governed by the fluid.

\subsection{Velocity Autocorrelations}

The impurity autocorrelation function satisfies
$dA/d\tau_q=-(1-q)A$. Therefore, its decay is similar to
(\ref{at-exp})
\begin{equation}
\label{at-imp} A(\tau_q',\tau_q)\sim A_0\exp[-(1-q)(\tau_q-\tau_q')]
\end{equation}
with $q$ replacing $p$ and $\tau_q$ replacing $\tau\equiv\tau_p$. For model
$A$ where the collision counters equal time, the decay remains
exponential. However, the impurity autocorrelation decays with a
different rate than the fluid. For model B, one can show that the
algebraic decay (\ref{at-1d}) holds asymptotically with $p$
replaced by $q$.  We conclude that while one-point velocity
statistics of the impurity are governed by the fluid, two-time
statistics are different.

\section{Mixtures}

Granular media typically consists of mixtures of granular particles of
several types.  In contrast with the impurity problem, the different
components of a mixtures are coupled to each other. Moreover, there are
numerous parameters, making mixtures much harder to treat analytically.
Remarkably, the impurity solution can be generalized to arbitrary
mixtures\footnote{Mixtures were treated analytically in the elastic Maxwell
  model \cite{ron}}. We detail the freely cooling case in one dimension.

Consider a binary mixture where particles of type 1 have mass $m_1$ and
concentration $c_1$, and similarly for particles of type $2$. We set
$c_1+c_2=1$.  Also, unit collision rates are considered for simplicity.
Collisions between a particle of type $i=1,2$ and a particle of type $j=1,2$
are characterized by the collision parameter $p_{ij}=(m_i-r_{ij})/(m_i+m_j)$,
with $r_{ij}$ the restitution coefficient. We denote the normalized velocity
distribution of component $i$ by $P_i(v,t)$ and its Fourier transform by
$F_i(k,t)$. The governing equations now couple the two distributions
\begin{eqnarray}
\label{fkt-mix} {\partial \over \partial t}\,F_i(k)+F_i(k) =
\sum_{j=1}^{2}c_j F_j\left(k-p_{ij}k\right)\,F_i\left(p_{ij}k\right).
\end{eqnarray}

Let $T_i=\int dv\,v^2\,P_i(v,t)$ be the temperature of the $i$th
component.  Writing $F_i(k,t)\cong 1-{1\over 2}k^2T_i$, the
temperatures evolve according
\begin{equation}
\label{temp-mix} {d\over d t}\left(\matrix{T_1\cr T_2}\right)=-
\left(\matrix{\lambda_{11}&\lambda_{12}\cr
\lambda_{21}&\lambda_{22} }\right)\, \left(\matrix{T_1\cr
T_2}\right).\qquad
\end{equation}
The diagonal matrix elements are
$\lambda_{ii}=2c_ip_{ii}(1-p_{ii})+c_j(1-p^2_{ij})$ with $j\ne i$,
and the off-diagonal matrix elements are
$\lambda_{ij}=-c_j(1-p_{ij})^2$. Therefore, the temperatures are
sums of two exponential terms, $\exp(-\lambda_{\pm}t)$. The decay
coefficients are the two (positive) eigenvalues
\begin{equation}
\lambda_{\pm}={1\over 2}\left[\lambda_{11}+\lambda_{22}
\pm\sqrt{(\lambda_{11}-\lambda_{22})^2+4\lambda_{12}\lambda_{21}}\,\right].
\end{equation}
Asymptotically, the term with the smaller decay rate
$\lambda\equiv \lambda_-$ dominates
\begin{equation}
T_i(t)\simeq C_ie^{-\lambda t},\qquad {\rm with} \qquad
C_i={(\lambda_+-\lambda_{ii})T_i(0)-\lambda_{ij}T_j(0)\over
\lambda_+-\lambda_-}.
\end{equation}
The ratio between the two temperatures approaches
$C_2/C_1=(\lambda-\lambda_{11})/\lambda_{12}$, and the two
components have different temperatures \cite{mp}. Furthermore, as
long as the components are coupled, both temperatures are finite.
For example, a vanishing $C_2$ implies that one of the
concentrations vanishes as $\lambda=\lambda_{1,1}$.

As in the cases of homogeneous gases and impurities, we seek a
scaling solution of the form $P_i(v,t)=e^{\lambda t/2}\,{\cal
P}_i(w)$ with $w=v\,e^{\lambda t/2}$, or equivalently,
$F_i(k,t)=f_i(z)$ with $z=|k|\,e^{-\lambda t/2}$. The scaling
functions $f_i(z)$ are coupled via the non-local differential
equations
\begin{eqnarray}
-{1\over 2}\lambda z f_i(z)+f_i(z) =\sum_{j=1}^2
c_jf_j(z-p_{ij}z)\,f_i(p_{ij}z).
\end{eqnarray}
Substituting the series solution (\ref{four-scl-exp-imp}) ,
$f_i(z)=\sum_{n=0}^\infty A_{i,n} z^ne^{-z}$, yields recursion
relations for the coefficients $A_{i,n}$
\begin{eqnarray}
\left(1-{n\lambda\over 2}\right)A_{i,n}+{\lambda\over 2}A_{i,n-1}
=\sum_{j=1}^2\sum_{m=0}^n c_jA_{j,m}A_{i,n-m}
(1-p_{ij})^mp_{ij}^{n-m}.
\end{eqnarray}
The small $z$ behavior $f_i(z)=1-{1\over 2}C_iz^2$ implies the first
three coefficients $A_{i,0}=A_{i,1}=1$, and $A_{i,2}=(1-C_i)/2$.  For
$n\ge 3$, the coefficients $(A_{1,n},A_{2,n})$ are solved recursively
in pairs. Each such pair satisfies two inhomogeneous linear
equations. Therefore, as in the impurity case, the velocity
distribution is an infinite series of powers of Lorentzians
(\ref{qw}). Although the two velocity distributions are different,
they have the same extremal behavior, ${\cal P}_i(w)\sim w^{-4}$.  It
is straightforward to generalize the above to mixtures with arbitrary
number of components, and to incorporate different collision rates, 
in particular, $K_{ij}=\sqrt{(T_i+T_j)/2}$.

\section{Lattice Gases}

Inelastic collisions generate spatial correlations and consequently,
inelastic gases exhibit spatial structures such as shocks. Thus far,
we considered only mean-field collision processes where there is no
underlying spatial structure. Random collision processes can be
naturally generalized by placing particles on lattice sites and
allowing only nearest neighbors to collide.

Consider a one-dimensional lattice where each site is occupied by a
single particle. Let $v_j$ be the velocity of the particle at site
$j$. The velocity of such a particle changes according to
Eq.~(\ref{rule-1d}) due to interactions with either of its two
neighbors. Time is conveniently characterized by the
collision counter $\tau$. In an infinitesimal time interval $d\tau$,
the velocity of a particle changes as follows
\begin{equation}
\label{rule} v_j(\tau+d\tau)=\cases{v_j(\tau)
&prob.~\,\,$1-2d\tau$;\cr
v_j(\tau)-(1-p)\left[v_j(\tau)-v_{j-1}(\tau)\right]  &prob.~\,\,
$d\tau$;\cr v_j(\tau)-(1-p)\left[v_j(\tau)-v_{j+1}(\tau)\right]
&prob.~\,\, $d\tau$.\cr}
\end{equation}
This process is stochastic and we are interested in averages over
all possible realizations of the process, denoted by an overline.
We consider random initial conditions where the average velocity
vanishes and no correlations are present: $\overline{v_j(0)}=0$
and $\overline{v_i(0)v_j(0)}=T_0\delta_{i,j}$.

Spatial velocity correlations satisfy closed equations as in the
Ising-Glauber spin model \cite{glauber}. For example, from the
dynamical rules (\ref{rule}), the average velocity
$V_j(\tau)=\overline{v_j(\tau)}$, obeys a discrete diffusion
equation
\begin{equation}
\label{diff} {d V_j\over d\tau}=(1-p)(V_{j-1}-2V_j+V_{j+1}).
\end{equation}
{}From the initial conditions, $V_j(0)=0$, we obtain $V_j(t)=0$.
Consider the spatial correlation function $\overline{v_iv_j}$. The
initial state is translationally invariant, so this property
persists. The correlation functions $R_n=\overline{v_jv_{j+n}}$
satisfy
\begin{eqnarray}
\label{R}
{d R_n\over d\tau}&=&2(1-p)(R_{n-1}-2R_n+R_{n+1}), \qquad n\geq 2;\nonumber\\
{d R_1\over d\tau}&=&2(1-p)\left[pR_0-(1+p)R_1+R_2\right];\\
{d R_0\over d\tau}&=&4p(1-p)\left[R_1-R_0\right].\nonumber
\end{eqnarray}
The initial conditions are $R_n(0)=T_0\delta_{n,0}$. Since we are
interested in the asymptotic behavior, we employ the continuum
approximation. The correlation function satisfies the diffusion
equation $\partial R/\partial \tau=2(1-p)\partial^2R/\partial^2n$,
so the solution is the Gaussian
\begin{eqnarray}
\label{pnt} R_n(\tau)\simeq {T_0\over\sqrt{8(1-p)\pi \tau}}\,\,
\exp\left[-\frac{n^2}{8(1-p)\tau}\right].
\end{eqnarray}
The temperature, $T\equiv R_0$ decays as $T(\tau)\simeq
T_0\,[8(1-p)\pi \tau]^{-1/2}$ in the long time limit\footnote{An
exact solution of the discrete equations (\ref{R}) is possible. It
yields an identical asymptotic expression for the temperature.}.

Although no correlations were present in the initial conditions,
spatial correlations develop at later times. The corresponding
correlation length $\xi$ grows diffusively with the collision counter,
$\xi\sim \tau^{1/2}$. The system consists of a network of domains of
typical size $\xi$. Inside a domain, velocities are strongly
correlated, and momentum conservation yields a relation between the
average velocity and the domain size, $v_*\sim \xi^{-1/2}\sim
\tau^{-1/4}$. This scale is consistent with the temperature behavior
above, $T\sim v_*^2$. In arbitrary dimension, this scaling argument
yields $T\sim \tau^{-d/2}$ \cite{be,bmp}. At least for scalar
velocities, the correlation functions obey closed equations in
arbitrary dimension \cite{bfk}, and this behavior can be also obtained
analytically.

The actual time dependence is determined from the collision rate.
We consider two choices.  In model A, the rate is proportional to
the typical velocity $K^2=T$.  In the physically more realistic
model B, the rate is proportional to the average relative velocity
$K^2=\overline{(v_{j+1}-v_j)^2}=2(R_0-R_1)\propto -dT/d\tau$. Hence
$K\sim \tau^{-\alpha}$ with $\alpha=d/4$ (model A) and
$\alpha=(d+2)/4$ (model B). The time $t=\int_0^\tau
d\tau'/K(\tau')$ grows as $t\sim \tau^{1+\alpha}$, and therefore,
\begin{equation}
T\sim t^{-\gamma} \qquad {\rm with}\qquad \gamma= \cases{{2d\over
d+4}&model A;\cr{2d\over d+6}&model B.}
\end{equation}
In either case, Haff's cooling law is recovered only in the
infinite dimension limit, $d\to\infty$. Otherwise, the appearance
of spatial correlations slows down the temperature decay.  As
neighboring particles become correlated their relative velocity
and consequently collision rate is reduced.  Qualitatively similar
behavior was shown for freely cooling inelastic gases,
namely, breakdown of the mean-field cooling law due to the formation of
strong spatial correlations among particles velocities
\cite{bcdr,nbc1}.

\section{Conclusions}

We presented analytic results for random inelastic collision
processes. In general, when the collision rate is uniform, the
convolution structure of the collision integrals translates to
products in Fourier space. While the governing equations are both
nonlinear and nonlocal, they are closed and amenable to analytical
treatment.  In the freely cooling case, a small wave number
analysis of the Fourier transform displays both regular and
singular terms. The regular terms yield the low order moments,
while the leading singular term gives the high-energy tail.
In the forced case, the Fourier transform is
an analytic function and complex residue analysis yields the
high-energy tail.

In the freely cooling case, the velocity distribution approaches a
scaling form and displays an algebraic large velocity tail.  In one
dimension, the scaling function is a universal; otherwise, it
depends on the degree of inelasticity. The exponent governing the
high-energy tail is typically very large, and may be difficult to
measure in practice.  We have also shown that sufficiently large
moments exhibit multiscaling behavior, and hence, are not
characterized by the temperature. The autocorrelation function
decays algebraically, with an exponent that depends on the
collision parameters. The spread of a tagged particle exhibits a
universal $\sqrt{\ln t}$ growth.

We demonstrated that an impurity immersed in a uniform fluid may or
may not mimic the background. Fixing the collision parameters, for
sufficiently low impurity mass, the impurity's velocity distribution,
velocity moments and extremal velocity statistics are all governed by
the fluid. However, there is a series of phase transitions occurring
at a series of increasing masses, where impurity moments of decreasing
orders decouple from the fluid.  These transitions indicate that
sufficiently heavy impurities are very energetic and effectively, they
experience a static fluid background.

For binary mixtures, we examined only primary velocity statistics and
showed that qualitatively, all components have the same temperature
decay and extremal velocity statistics. It remains to be seen whether
the different components may exhibit different asymptotic behaviors of
more subtle velocity statistics such as the high order moments. The
impurity case demonstrates how in certain limiting cases, mixtures may
display anomalous behavior.

In the forced case, injection of energy counters the energy
dissipation and the system relaxes toward a steady state. We
considered white noise forcing, and showed that the steady state
distribution has an exponential high-energy tail. Steady state
characteristics in the forced case are directly related to time
dependent relaxation characteristics in the freely cooling case.

Results obtained in the framework of the inelastic Maxwell model are
exact for random collision processes where spatial structure is
absent, and particles collide irrespective of their relative
velocities. Such results are an uncontrolled approximation of the
inelastic hard sphere problem. Nevertheless, as a conceptual tool, the
inelastic Maxwell model is powerful. For example, it demonstrates the
development of correlations among the velocity components, as well as
spatial correlations.  It also raises doubts concerning the
suitability of several widely used techniques such as perturbation
expansions in the quasi-elastic limit, and expansions in terms of
Sonine polynomials.

We demonstrated that a lattice gas generalization of the Maxwell model
remains tractable when the kinematic constraint (particles should
collide only if they are moving toward each other) is ignored.  Taking
this constraint into account leads to shock-like structures in one
dimension and to vortices in two dimensions \cite{bmp}. Lattice models
can therefore be used to study development of spatial correlations and
spatial structures in inelastic gases.

We presented in detail the most basic realization of the Maxwell
model. Several other generalizations are feasible. Experiments can
now measure the distributions of impact angles and of the
effective restitution coefficients. Such phenomenological
information can be incorporated into the rate equations. The
integration measure can be redefined to give different weights to
different angles. Moreover, a distribution of restitution
coefficients can be introduced by integrating the collision
integrals with respect to the collision parameters. For example,
random collision processes were successfully used to model the
role of the boundary in driven gases \cite{vm}.

We restricted our attention to the case where all moments of the
initial velocity distribution are finite. However, there are other
infinite energy solutions of the Boltzmann equation. For example,
any Lorentzian, $F(z)=\exp(-C z)$, is a steady state solution of
Eq.~(\ref{four-eq-1d}). A remaining challenge is to classify the
evolution of an arbitrary velocity distribution. Also, it will be
interesting to characterize the full spectrum of extremal
behaviors, characterizing velocities far larger than the typical
velocity.

Finally, to model transport and other hydrodynamics problems one has to
incorporate the spatial dependence explicitly.  In the framework of the
Maxwell model the simplest such problem -- the shear flow -- has been
recently investigated by Cercignani \cite{carlo}. For inelastic hard spheres,
a number of unidirectional hydrodynamic flows were analyzed by Goldshtein
and co-workers \cite{sasha1,sasha2,sasha3}. It would be interesting to
investigate these problems in the framework of the Maxwell model.

\subsubsection{Acknowledgments} We have benefited from discussions and
correspondence with A.~Baldassarri and M.~H.~Ernst.  We also thank
N.~Brilliantov, A.~Goldshtein, A.~Puglisi, S.~Redner, and
H.~A.~Rose for useful discussions. This research was supported by
DOE (W-7405-ENG-36) and NSF(DMR9978902).

\end{document}